\documentclass[12pt]{article}
\usepackage{a4wide,epsfig,amsfonts,latexsym}

\newcommand{\calB}{{\cal B}}
\newcommand{\calT}{{\cal T}}

\newcommand{\braced}[1]{{ \left\{ #1 \right\} }}
\newcommand{\angled}[1]{{ \left\langle #1 \right\rangle }}

\newcommand{\mymod}[1]{{ \,(\mbox{\rm mod}\, #1)}}

\newcommand{\bfe}{{\mathbf e}}
\newcommand{\bfb}{{\mathbf b}}
\newcommand{\NP}{\mbox{$\mathbb{NP}$}}

\newtheorem{theorem}{Theorem}
\newtheorem{fact}{Fact}
\newenvironment{proof}
	{\begin{trivlist}\item[]{\em Proof.}}
	{\hspace*{\fill}$\Box$\end{trivlist}}

\begin{document}

\title{A Note on Tiling under Tomographic Constraints}

\author{Marek Chrobak\thanks{Department of Computer Science,
			University of California, Riverside,
			CA 92521.
			Research supported by NSF grant CCR-9988360.
			Email: {\tt marek@cs.ucr.edu}.}
	\and
	Peter Couperus\thanks{Department of Mathematics, University of
			Washington, Seattle, WA 98195-4350.
			Email: {\tt couperus@math.washington.edu}.}
	\and
	Christoph D\"urr\thanks{LRI, Universit\'e Paris-Sud,
			B\^atiment 490, F-91405 Orsay c\'edex, France.
			Email: {\tt durr@lri.fr}.}
	\and
	Gerhard Woeginger\thanks{Department of Mathematics, 
	University of Twente, P.O. Box 217, 7500 AE Enschede, 
	The Netherlands. Email: {\tt g.j.woeginger@math.utwente.nl}.
	This work was conducted while the 
	fourth author was affiliated with the Institut fuer Mathematik, 
	TU Graz, Austria.}
	}

\maketitle

\begin{abstract}
Given a tiling of a 2D grid with several types of tiles, we can count
for every row and column how many tiles of each type it intersects.
These numbers are called the \emph{projections}. We are interested in
the problem of reconstructing a tiling which has given projections.
Some simple variants of this problem, involving tiles that are
$1\times 1$ or $1\times 2$ rectangles, have been studied in the past,
and were proved to be either solvable in polynomial time or
{\NP}-complete. In this note we make progress toward a comprehensive
classification of various tiling reconstruction problems, by proving
{\NP}-completeness results for several sets of tiles.
\end{abstract}


\section{Introduction}

In \emph{Discrete Tomography} we want to reconstruct a discrete object
from its projections.  This paper is concerned with the reconstruction
of \emph{tilings}. We are given a collection of \emph{tiles}, where
each tile can have a different shape.  A tiling is a placement of
non-overlapping copies of the tiles on a $n\times n$ grid, where each
copy is obtained by translating one of the tiles.  (In this note we do
not allow tile rotations, although one could also consider the variant
with rotations.)  \emph{Projections} of a tiling determine the number
of tiles intersected by each row and column. Given such projections, we
wish to reconstruct a tiling consistent with these projections, or to
report that such a tiling does not exist.

Formally, a tile $t$ is defined to be a finite subset of $\mathbb
Z^2$.  In this paper we only consider tiles that are hole-less
polyominoes.  By $(i,j)+t=\{(i+i',j+j'):(i',j')\in t\}$ we denote the
translation of $t$ by vector $(i,j)\in \mathbb Z^2$.  Fix a finite
multiset of tiles $\calT = \{t_1, t_2, \ldots, t_h\}$.  Without loss
of generality we assume that every tile $t_k$ contains $(0,0)$, the
so-called \emph{center of the tile}, and in this paper it will always
be the upper-left corner.  We refer to the index $k$ as the
\emph{type} of the tile. The tiles are identified by their type, and
different tiles may have the same shape.  One can think of tiles which
are of the same shape but of different types as being of different
colors.


\begin{figure}[tbh]
\centerline{\epsfig{width=9em,file=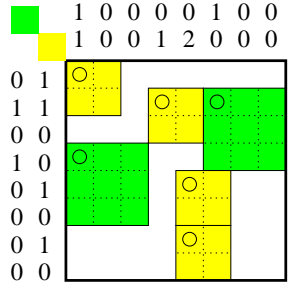}}
\caption{A tiling of the $8\times 8$ grid with its projections.  The
tile centers are marked by a circle.}
\label{proj}
\end{figure}


A \emph{$\calT$-tiling} of a grid $G = \mathbb Z_n\times \mathbb Z_n$
is a finite set $T\subseteq \mathbb Z_n\times \mathbb Z_n \times
[1,h]$, such that the sets $(i,j)+t_k$, for all $(i,j,k)\in T$,
are disjoint and contained in $G$.  If $\calT=\braced{t}$, we will
sometimes write simply \emph{$t$-tiling} instead of $\braced{t}$-tiling.
The \emph{center projections} of $T$ are vectors
$r, c\in \mathbb N^{n\times h}$, where
\begin{eqnarray*}
	r_{i,k} &=& |\{ (i,j,k) \in T \}| \quad\textrm{and} \\
	c_{j,k} &=& |\{ (i,j,k) \in T \}|.
\end{eqnarray*}
The numbers $r_{i,k}$ count the number of tiles of type $k$ whose
center is in row $i$, the numbers $c_{j,k}$ count the same for column
$j$.  In a similar manner we define the \emph{cell projections}
$r',c'$ of $T$, where we count for each row, each column, and each type
the number of cells covered by that type of tile.

If tilings $T$ and $T'$ are disjoint, then projections of $T\cup T'$
are the sums of the projections of $T$ and of $T'$.  (This is true for
both types of projections.)  Therefore the set of projections of all
tilings $T$ with a single tile ($|T|=1$) spans the set of all
projections.  The canonical bijection between \emph{single-tile center
projections} and \emph{single-tile cell projections} implies a
bijection between all center and cell projections.  From now on we
will use the term ``projection" for center projections, unless stated
explicitly otherwise.

Note that we do not require the tilings to cover the whole grid.
Tilings that cover the whole grid are called \emph{complete}.  Each
tiling problem can be mapped into an equivalent complete tiling by
adding one ``clear'' $1\times 1$ square tile whose row cell projections
are $n$ minus the total sums of the other tiles' row cell projections,
and the column projections are defined analogously.

Figure~\ref{proj} illustrates this definition for the tile set
$\calT = \{t_1,t_2\}$, where $t_1$ is the $3\times 3$ square and
$t_2$ is the
$2\times 2$ square.  The numbers on the left are the projections
$r_{i,k}$ and the numbers on the top are the $c_{j,k}$.  Columns are
numbered from left to right and rows from top to bottom, with indices
ranging from $0$ to $7$.  For example $c_{4,1}=2$ because column
$4$ contains two centers of tile $t_1$.

Given a tiling $T$, the computation of its projections is
straightforward.  Consider now the inverse problem: given the vectors
$r,c$, find a $\calT$-tiling $T$ with projections $r,c$.  This
is called the \emph{$\calT$-reconstruction problem}.  The related decision
problem (``is there such a tiling $T$?'') is called the
\emph{$\calT$-consistency problem}, or simply the \emph{$\calT$-tiling
problem}.


\paragraph{Our results.}
For some types of tiles the reconstruction problem is easy to solve,
while for other it may be hard.
Table~\ref{table} summarizes the complexity of various tiling 
reconstruction problems, including both our results and previous work.
In this table, by ``{\NP}-complete" we mean 
that the consistency problem is {\NP}-complete.


\begin{table}[tb]
\centerline{
\begin{tabular}{lll}
type of tiles 		& complexity 			& reference
\\\hline\\[-1em] 
$\left\{\begin{array}{c}
  \epsfig{width=1em,file=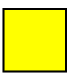}
	\end{array}
\right\}$
			& $O(n^2)$ &\cite{Ryser63}
\\[1em]
$\left\{\begin{array}{c}
  \epsfig{width=1em,file=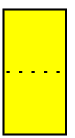}
	\end{array}
\right\}$
			& $O(n^2)$ &\cite{DGRR00,Picouleau99}
\\[1em]
$\left\{\begin{array}{c}
  \epsfig{width=2em,file=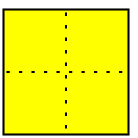}
	 \end{array}
\right\}$
			& $\ge$ ``2-atom problem'' 
					& Theorem~\ref{thm-1square}
\\[1em]
$\left\{\begin{array}{c}
  \epsfig{width=2em,file=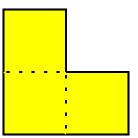}
	 \end{array}
\right\}$
			& {\NP}-complete
						& Theorem~\ref{thm-Lshape}
\\[1em]
$\left\{\begin{array}{c}
  \epsfig{width=4em,file=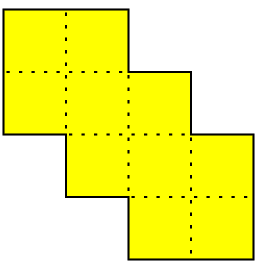}
	 \end{array}
\right\}$
			& {\NP}-complete &Theorem~\ref{thm-shape}
\\[1em]\hline\\[-1em] 
$\left\{\begin{array}{c}
  \epsfig{width=1em,file=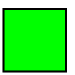},
  \epsfig{width=1em,file=y1x1.eps}
	\end{array}
\right\}$
			& open & ``2-atom problem''
\\[1em]
$\left\{\begin{array}{c}
  \epsfig{width=2em,file=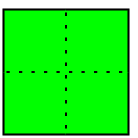},
  \epsfig{width=2em,file=y2x2.eps}
	\end{array}
\right\}$
			& {\NP}-complete & Theorem~\ref{thm-2squares}
\\[1em]
 $\left\{\begin{array}{c}
   \epsfig{width=1em,file=g1x1.eps},
   \epsfig{width=1em,file=y2x1.eps}
 	\end{array}
 \right\}$
 			& $\ge$ ``2-atom problem'' & (obvious)
 \\[1em]
$\left\{\begin{array}{c}
  \epsfig{width=2em,file=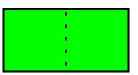},
  \epsfig{width=1em,file=y2x1.eps}
	\end{array}
\right\}$
			& {\NP}-complete & Theorem~\ref{thm-2domino}
\\[1em]
$\left\{\begin{array}{c}
  \epsfig{width=1em,file=g1x1.eps},
  \epsfig{width=2em,file=y2x2.eps}
	\end{array}
\right\}$
			& {\NP}-complete & Theorem~\ref{thm-square}
\\[1em]\hline\\[-1em] 
$\left\{\begin{array}{c}
  \epsfig{width=1em,file=g1x1.eps},
  \epsfig{width=1em,file=y1x1.eps},
  \epsfig{width=1em,file=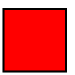}
	\end{array}
\right\}$
			& {\NP}-complete &\cite{ChrobakDurr01}
\\[1em]\hline\\[-1em] 
\end{tabular}
}
\caption{The complexity of different versions of the
reconstruction/consistency problem.}
\label{table}
\end{table}


\paragraph{The $l$-atom reconstruction problem.}
The simplest tile is a $1\times 1$ square, which we call a \emph{cell}
or an \emph{atom} (the original motivation for this problem came from
the reconstruction of polyatomic structures).  When $\calT$ consists
of $l$ different cells, we will refer to the $\calT$-tiling problem as
the \emph{ $l$-atom problem}. Reconstructing 1-atom tilings is easy
and can be solved in time linear in the size of the grid, as shown in
1957 by Ryser~\cite{Ryser57,Ryser63}. For 3 or more
atoms (cells of different type), the reconstruction problem is
{\NP}-hard~\cite{ChrobakDurr01} (see also \cite{GaGrPr00}). For 2
atoms, the complexity of the problem remains open.


\paragraph{One tile.}
For a single tile, it is known that if the tile is a horizontal
\emph{bar}, i.e.\ a rectangle of height one, the problem is as easy as
reconstructing 1-atom tilings~\cite{DGRR00,Picouleau99}.  There exist
other types of tiles, however, for which the problem is {\NP}-complete.
Two such tiles are given in Table~\ref{table}.  The problem remains
open for rectangular tiles, even for the $2\times 2$ square.


\paragraph{Two tiles.}
For pairs of tiles the situation is quite different.  For horizontal
and vertical dominoes --- $1\times2$ and $2\times1$ rectangles --- the
problem is {\NP}-hard.  However, the case when the domino tiling
is required to be complete is open.  The problem is also open for
vertical dominoes and single cells. For squares the problem is
{\NP}-hard, both for two types of $2\times2$ squares, and for a single
$2\times2$ square and a cell.


\section{{\NP}-hardness Proofs}

We now present our {\NP}-hardness results. In our proofs we reduce the
$3$-atom problem to the given version of the $\calT$-tiling problem. A
similar strategy was used earlier by D\"urr, Goles, Rapaport and
R\'emila~\cite{DGRR00} to show that reconstructing tilings of given
sub-grids --- grids with forbidden regions --- with only vertical and
horizontal dominoes is {\NP}-hard (even if the tilings are required to be
complete).

The general idea of the proofs can be summarized as follows.  We think
about the $3$-atom problem as a $4$-atom \emph{complete tiling}
problem, with an additional ``clear'' atom.  For convenience, we name
each possible tile in the 3-atom problem as \emph{yellow},
\emph{blue}, \emph{red} or \emph{clear}.  Throughout this section, by
$\angled{r, c}$ we will denote the given instance of the 3-atom
problem.  We will map $\angled{r,c}$ into an instance $\angled{r',c'}$
of the $\calT$-tiling problem under consideration.  In all proofs we
assume, without loss of generality, that $\sum_i r_{i,k}= \sum_j c_{j,k}$
for all $k$. This assumption is valid, since we can
extend any mapping to instances in which
$\sum_i r_{i,k}\neq \sum_j c_{j,k}$, by mapping them
into an arbitrary fixed negative instance $\angled{r',c'}$,
say to one in which the totals of row sums are not equal
to the totals of column sums. This does not affect the
asymptotic running time nor the correctness of the transformation.

For simplicity, assume first that $\calT$ contains just one tile.
To construct $\angled{r',c'}$, we choose a small $d\times d$ grid
$\calB$, called a \emph{block}, that can be tiled in only four
possible ways (this restriction will be relaxed in some proofs).  Each
of these four so-called \emph{admissible} block tilings will
correspond to one atom. Instance $\angled{r',c'}$ will have grid
dimensions $nd\times nd$. We view this grid as an $n\times n$ matrix
consisting of $d\times d$ blocks.  A segment of rows numbered
$id,\dots,(i+1)d-1$ will be referred to as a \emph{block-row} $i$. The
transformation maps $r_{i,1},\dots,r_{i,4}$, that is, the atom
projections of row $i$, into a length-$d$ vector which is a projection
of block-row $i$.  This vector is the linear combination of the
projections of the admissible tilings of $\calB$ with coefficients
$r_{i,1},\dots,r_{i,4}$. The column projections are mapped in the same
way.

If $\calT$ has $h>1$ tiles, the transformation is the same,
except that now the block projections are not length-$d$ vectors but
$d\times h$ matrices.

Obviously, for any tile set $\calT$, the $\calT$-consistency problem
is in {\NP}. For any choice of $\calB$ and its admissible tilings,
the method outlined above can be implemented in polynomial
time. It also has the property that if $\angled{r,c}$
has a solution then so does $\angled{r',c'}$. For if
$T$ is a solution of the 3-atom problem with
projections $\angled{r,c}$, then the tiling $T'$ obtained by replacing
each atom by its corresponding admissible block is a
$\calT$-tiling with projections $\angled{r',c'}$.
Thus the above ingredients
of {\NP}-completeness arguments will be omitted in the proofs
below, and we will focus exclusively on proving the following
implication: if $\angled{r',c'}$ 
has a solution then $\angled{r,c}$ has a solution.

The main difficulty is to construct $\calB$ to make
this latest implication work. In other words, we need the property 
that any tiling of the resulting instance $\angled{r',c'}$ can be
transformed back into a solution of $\angled{r,c}$. To achieve this,
we choose $\calB$ and the admissible tilings so that
the following two conditions hold: 
\begin{description}
\item{(npc1)} any tiling with projections $\angled{r',c'}$
consists only of admissible block tilings, and
\item{(npc2)}
from the projections of the block-rows we can
uniquely extract the projections of the atoms in 
the corresponding rows of the 3-atom problem.
\end{description}
To enforce condition (npc1), we use techniques inspired by classical
structure theorems for realizations of $0-1$ matrices with given
projections \cite{Ryser63}.  Another useful method involves the total
counts of different colors. By $Y$, $B$, $R$ and $C$
we will denote the total number of yellow, blue, red and clear atoms
in $\angled{r,c}$. We have $B+Y+R+C = n^2$.  The block projections
impose additional restrictions on how many centers of the tiles in
$\calT$ can occur on certain positions in the blocks. These
restrictions can be expressed in terms of numbers $Y$, $B$, $R$ and
$C$.  By investigating these constraints, we prove that non-admissible
tilings cannot occur.

We now discuss condition (npc2). Suppose that there is a $\calT$-tiling
$T'$ with projections $\angled{r',c'}$. By (npc1),
each block in $T'$ is admissible. We transform $T'$ into
a solution $T$ of the 3-atom problem by replacing each admissible
block by its corresponding atom. To satisfy (npc2),
we need to show that the projections of $T$ are $\angled{r,c}$.

Number the admissible tilings from $1$ to $4$ and
name their row projections $\bfe_1,\bfe_2,\bfe_3,\bfe_4$.
Let $\bfb_i$ be the projection of block-row $i$ in $T'$
and $q_j$ the number of
blocks in block-row $i$ with the $j$th admissible tiling.
Then the numbers $q_j$ satisfy:
\begin{eqnarray}
        q_1\bfe_1 + q_2\bfe_2 + q_3 \bfe_3 + q_4 \bfe_4 &=& \bfb_i.
                \label{eqn: proj equation}
\end{eqnarray}
By the construction, equation (\ref{eqn: proj equation}) has
a solution $q_j = r_{i,j}$, for $j = 1,2,3,4$. 
For (npc2) to hold, we need to ensure that
(\ref{eqn: proj equation}) does not have any other
solutions in non-negative integers.  This can be
easily accomplished by choosing the admissible tilings for which the
projections $\bfe_1,\bfe_2,\bfe_3,\bfe_4$ are linearly independent. 

In fact (npc2) will hold even for a weaker condition.  Note that the
numbers $r_{i,j}$ satisfy $r_{i,1}+r_{i,2}+r_{i,3}+r_{i,4} = n$.  So
we extend each $\bfe_j$ by adding to it one coordinate with value $1$,
and we similarly extend each vector $\bfb_i$ by adding to it one
coordinate with value $n$.  (Technically, the $\bfe_j$ and $\bfb_i$
can be $d\times h$ matrices, but for the purpose of the transformation
we can as well treat them as vectors of length $dh$. Then the extended
vectors will have length $dh+1$.) If these new vectors
$\bfe_1,\bfe_2,\bfe_3,\bfe_4$ are linearly independent, we can use our
admissible tilings for the transformation. Although we do not use it
in the paper, it may be worth to mention that the linear independence
of these extended vectors is equivalent to a condition called affine
linear independence (see page 3 of~\cite{GrLoSc}).


\begin{theorem}					\label{thm-shape}
Let $t$ be the tile \epsfig{width=3em,file=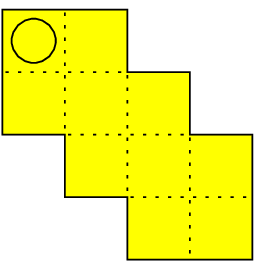}.  The
consistency problem for $t$ is {\NP}-complete.
\end{theorem}

\begin{proof}
The proof follows the method outlined above. We reduce the 3-atom
consistency problem to the $t$-tiling.  We treat the 3-atom problems as
the (equivalent) complete $4$-atom  problem, by adding an extra ``clear''
cell tile.  We use a block of size $7\times 7$.  The admissible
tilings are the tilings of the $7\times 7$ block with three tiles
$t$. There are exactly four admissible tilings. These tilings and their
associations to different atoms are shown in Figure~\ref{shape}.  Using
the projections of these admissible block tilings, we map any instance
$r,c\in\mathbb N^{n\times 4}$ of the 3-atom consistency problem into
an instance $r',c'\in\mathbb N^{7n}$ of the $t$-tiling problem.


\begin{figure}[hbt]
\centerline{\epsfig{width=35em,file=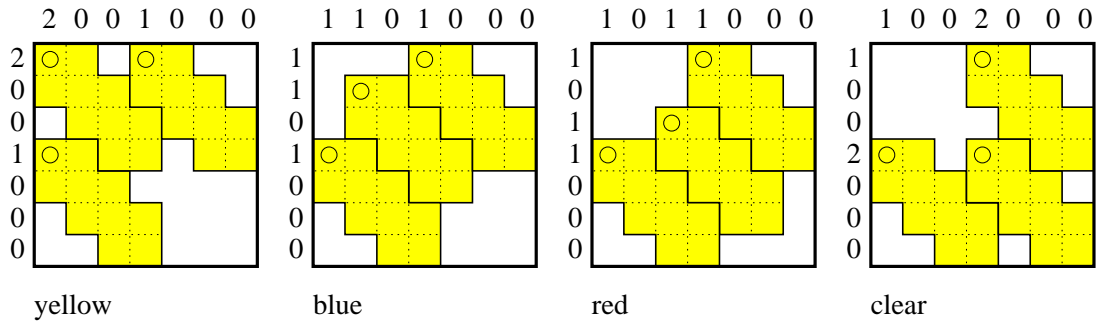}}
\caption{The four admissible tilings of a block with 3 tiles $t$.}
							\label{shape}
\end{figure}


The transformation works as follows.
Let $\bfe_1, \bfe_2, \bfe_3, \bfe_4 \in \mathbb
N^{7}$ denote the row projection vectors of the four tilings
in Figure~\ref{shape}.
For every row $i$,  its projections form a 4-dimensional 
vector $(r_{i,1}, r_{i,2}, r_{i,3}, r_{i,4})$.
We will map it into $\mathbb N^{7}$.
The projections $r'$ in the resulting instance are defined by
\[
	\left( \begin{array}{c}
	r'_{7i  } \\
	r'_{7i+1} \\
	r'_{7i+2} \\
	r'_{7i+3} \\
	r'_{7i+4} \\
	r'_{7i+5} \\
	r'_{7i+6}
	       \end{array} \right)
	\;=\;
			r_{i,1} \bfe_1 + 
			r_{i,2} \bfe_2 +
			r_{i,3} \bfe_3 + 
			r_{i,4} \bfe_4.
\]
In other words we set $r'_{7i}=r_{i,1}+n$, $r'_{7i+1}=r_{i,2}$,
$r'_{7i+2}=r_{i,3}$, $r'_{7i+3}=r_{i,4}+n$ and $r'_{7i+l}=0$ for
$l = 4,5,6$. The column projections $c'$ are determined in a similar
manner.

We need to show that if $\langle r',c'\rangle$ has a solution then
$\langle r,c\rangle$ also has a solution.  To this end,
let $T$ be an arbitrary solution to $\langle r',c'\rangle$.  We claim
that in $T$ every block is in one of the four configurations of
Figure~\ref{shape}.  This is true because the rows and columns whose
indices modulo 7 are greater than 3 have projection 0 and therefore all
tiles $t$ are strictly contained in a block. Further, $\langle
r',c'\rangle$ requires $3n^2$ tiles $t$ in total, and each of the
$n^2$ blocks contains at most three tiles of type $t$. 


\begin{figure}[htb]
\centerline{\epsfig{width=33em,file=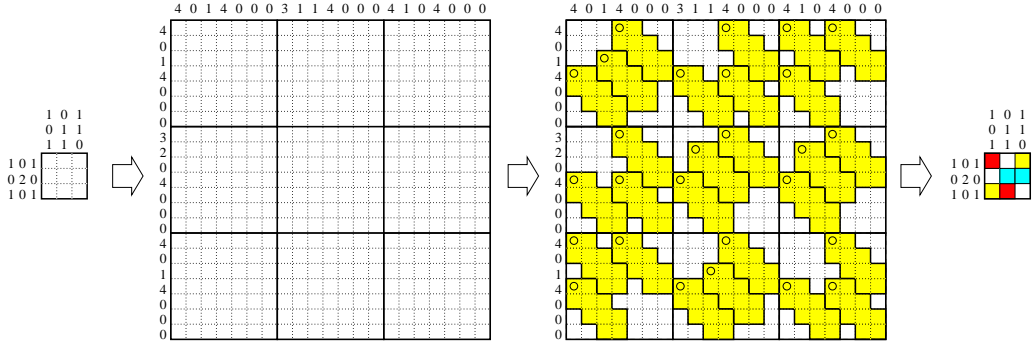}}
\caption{Reducing the 3-atom reconstruction problem to the $t$-tiling
reconstruction problem.}
\label{idea}
\end{figure}


The vectors $\bfe_k$ are linearly independent.  So the
projections $(r'_{7i+k})_{k\in\mathbb Z_7}$
of a block row $i$ can be uniquely written as $\sum_{k=1}^4 r_{i,k}
\bfe_k$.  This ensures that $T$ can be mapped into a 3-atom
tiling with projections $r,c$ (see Figure~\ref{idea}).
\end{proof}


For the next result we need the following classical result
on the structure of $0-1$ matrices with given projections \cite{Ryser63}.
We state this fact in terms of tiling with cells.

\begin{fact}						\label{fact}
Let $r,c\in \mathbb Z^n$ be an $n\times n$ instance of the tiling
problem with cells.  Let $I\subseteq \mathbb Z_n$ be a row set and
$J\subseteq \mathbb Z_n$ a column set.  If
\[
	\sum_{i\in I} r_i - \sum_{j\in \overline J} c_j = |I\times J|,
\]
then in every solution the set $I\times J$ must be completely tiled and
the set $\overline I \times \overline J$ must be completely empty.
\end{fact}

\begin{proof}
Let $T$ be a solution to $\langle r, c\rangle$.  Let $a$ be the number
of cells in $I\times J$, $b$ the number of cells in $I\times \overline
J$ and $c$ the number of cells in $\overline I \times \overline J$.
Then 
\[
	\sum_{i\in I} r_i - \sum_{j\in \overline J} c_j =
		(a+b)-(b+c)=a-c.
\]
If $a-c = |I\times J|$ then $a=|I\times J|$ and $c=0$, which concludes
the proof.
\end{proof}


\begin{theorem}					\label{thm-2domino}
The consistency problem for $1\times 2$ and $2\times 1$ rectangles
(dominoes) is {\NP}-complete.
\end{theorem}

\begin{proof}
We follow the idea outlined at the beginning of this section.
We use a $3\times 3$ block. The four admissible tilings of the
block are shown in Figure~\ref{2domino}. 


\begin{figure}[htb]
\centerline{\epsfig{width=30em,file=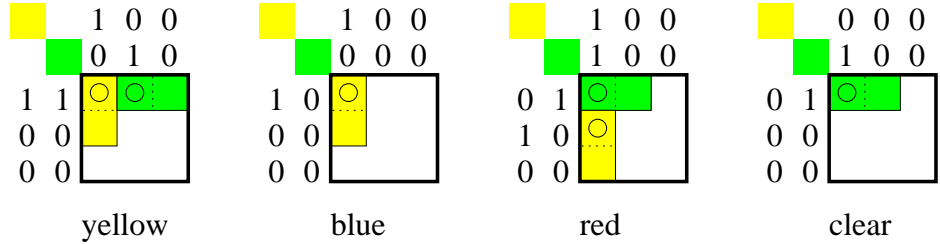}}
\caption{Four admissible tilings of the $3\times3$ grid with dominoes.}
						\label{2domino}
\end{figure}


We need to show that if $\angled{r',c'}$ has a solution then
$\angled{r,c}$ has a solution. Note that the row and column projection
matrices of the four tilings in Figure~\ref{2domino} are not linearly
independent, but at least satisfy the weaker condition described on
page~\pageref{thm-shape}, which is enough for the reduction.
Therefore to complete the proof we need to show that in any solution
of the tiling instance every block is admissible.

Let $I = J = \{ i\in \mathbb Z_{3n} : i \bmod 3 > 0\}$
be row and column sets. Denote the ``clear'' cell by $t_3$.
Recall that $Y,B,R$ and $C$ denote the total numbers of yellow,
blue, red and clear atoms in $\angled{r,c}$. Then we have
$\sum_{i\in I} r'_{i,3} - \sum_{j\in \overline J} c'_{j,3}
	= (5Y+5B+4R+6C) - (Y+B+2C)
	= 4Y+4B+4R+4C=4n^2$.
Fact~\ref{fact} implies that in every block dominoes can only
appear in the first row or first column of each block and the
top-left cell 
is always covered by a domino. All the tilings that satisfy these
condition are the admissible tilings in Figure~\ref{2domino}.
\end{proof}


\begin{theorem}					\label{thm-2squares}
The consistency problem for two types of $2\times2$ squares
is {\NP}-complete.
\end{theorem}

\begin{proof}
We refer to the two types of $2\times2$ squares as \emph{light}
and \emph{dark} squares.
We use the $4\times 4$ block and four admissible block tilings
shown in Figure~\ref{2squares}.
The row and cell projection matrices of the admissible
block tilings are linearly independent. Thus,
to complete the proof, we have to show that any solution of
$\angled{r',c'}$ uses only the four admissible block tilings.


\begin{figure}[htb]
\centerline{\epsfig{width=35em,file=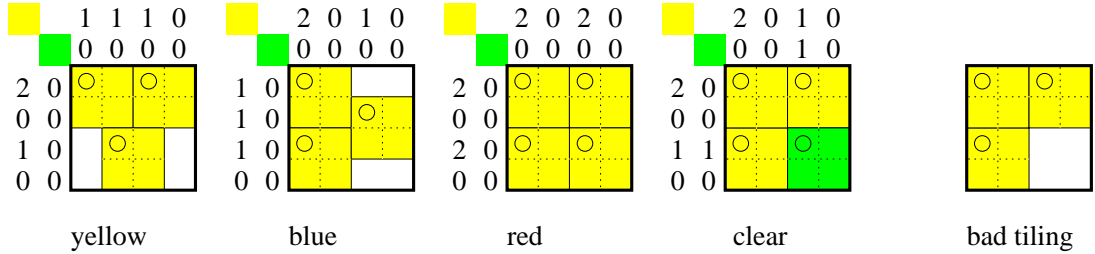}}
\caption{Four tilings of the $4\times4$ grid with two types of squares.}
						\label{2squares}
\end{figure}


If we consider the cell projections rather than the center
projections, we see that row $1$ and column $1$ of every block must be
completely tiled with light squares 
(recall that rows and columns are numbered from $0$).  
Therefore a block can only be in
one of the five tilings shown in Figure~\ref{2squares}.  The fifth
tiling --- which is called \emph{bad tiling} --- has the same row
projections as a ``yellow'' block and the same column projections as a
``blue'' block. By column projections for columns $j = 1\mymod{4}$,
in any tiling
there will be $Y$ ``yellow" blocks, and by row projections for rows $i
= 1\mymod{4}$, there will be $B$ ``blue" blocks. There are $C+R$
remaining blocks, and $4(C+R)$ tiles must appear in these remaining
blocks, so each of these remaining blocks must have four tiles. Thus
the bad tiling cannot occur.
\end{proof}


The same technique can be used to reduce the 2-atom problem to the
single-type square tiling problem.  In this reduction,
only the first 3 block tilings of Figure~\ref{2squares} are used.
It can also be used to
prove {\NP}-completeness of the cell-and-square tiling problem.
In this reduction, the dark square is replaced by a cell in
Figure~\ref{2squares}, without modifying the projections.


\begin{theorem}				\label{thm-1square}
If the consistency problem for $2\times 2$ squares can be solved
in polynomial time, then the $2$-atom problem can be solved in
polynomial time.
\end{theorem}


\begin{theorem}					\label{thm-square}
The consistency problem for $1\times1$ cells and $2\times 2$ squares
is {\NP}-complete.
\end{theorem}


\begin{theorem}					\label{thm-Lshape}
Let $t$ be the \emph{L-shaped} tile
\epsfig{width=2em,file=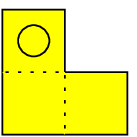}.  The consistency problem
for $t$ is {\NP}-complete.
\end{theorem}

\begin{proof}
We reduce the 3-atom problem to the $t$-tiling problem using the
admissible block tilings of Figure~\ref{Lshape}.  The first three
tilings correspond to the colored atoms, and the two last tilings (with
identical projections) correspond to the clear atom. The row and
column projections of the admissible tilings are linearly independent.


\begin{figure}[htb]
\centerline{\epsfig{width=35em,file=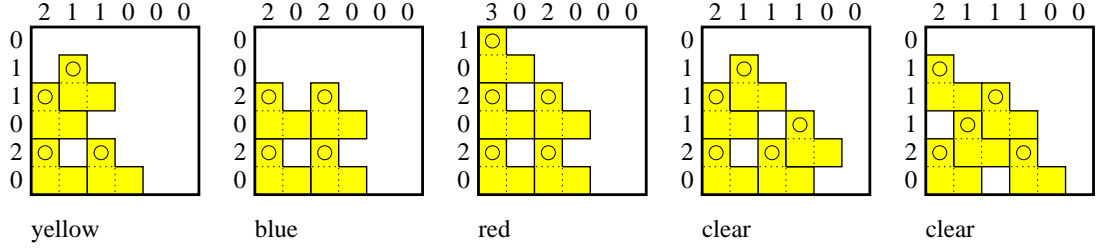}}
\caption{Five admissible tilings of the $6\times6$ block
with the L-shaped tile. The last two tilings have the same projections.}  
							\label{Lshape}
\end{figure}


It is sufficient to show that in any solution $T$ to $\angled{r',c'}$
every block must be admissible.  We define the matrix $M\in\mathbb
N^{6\times6}$ where $m_{i,j}$ is the number of tiles in $T$ whose
center equals $(i,j)$ modulo 6. Row and column sums of $T$ imply that
$M$ must have the following form (the numbers on the left and on top
are the row and column sums of $M$):
\[
	    \begin{array}{r|cccccc}
	    & 2n^2 +R & Y+C     & n^2+B+R & C  	    & 0	& 0      \\ \hline
	R   & m_{0,0} & m_{0,1} & m_{0,2} & m_{0,3} & 0 & 0\\
	Y+C & m_{1,0} & m_{1,1} & m_{1,2} & m_{1,3} & 0 & 0\\
    n^2+B+R & m_{2,0} & m_{2,1} & m_{2,2} & m_{2,3} & 0 & 0\\
	C   & m_{3,0} & m_{3,1} & m_{3,2} & m_{3,3} & 0 & 0\\
       2n^2 & m_{4,0} & m_{4,1} & m_{4,2} & m_{4,3} & 0 & 0\\
	0   & 0       & 0       & 0       & 0       & 0 & 0
	    \end{array}
\]
The tiles centered at $(i,j)$ and $(i-1,j-1)$ overlap, and they
both overlap with the tile centered at $(i-1,j)$ or $(i,j-1)$.
This introduces two additional constraints on the matrix: for every $i,j$
\begin{eqnarray}
	m_{i-1,j-1} + m_{i-1,j} + m_{i,j} &\le& n^2, \quad\mbox{\rm and }
						\label{eq-L1} \\
	m_{i-1,j-1} + m_{i,j-1} + m_{i,j} &\le& n^2,
						\label{eq-L2}
\end{eqnarray}
where we set $m_{i,j}=0$ for $i=-1$ or $j=-1$.

Every block must have $2$ centers in row $4$ (recall that rows and
columns are numbered from $0$).  So row $3$ can have at most $1$
center.  We now consider sub-blocks that consist of rows $3,4,5$.  By
the above argument, the tiles that are fully contained in these
sub-blocks must have one of the following configurations:

\bigskip

\centerline{\epsfig{width=5in,file=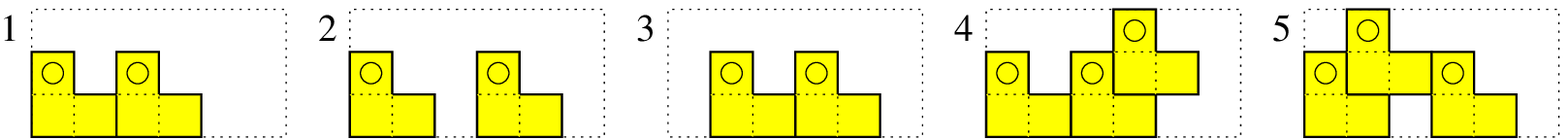}}

\bigskip

This immediately gives $m_{3,0} = m_{3,2} = 0$, $m_{4,0} = n$.
Let $a_s$ be the number of blocks whose last three rows
are in configuration
of type $s$ above, for $s=1,...,5$. Then $a_4 = m_{3,3}$ and
$a_2+a_3+a_5 = m_{4,3}$. Since the projections of row 3 and column 3
of $M$ are equal $C$, we get $a_4 + a_5 = C$ and $a_2+a_3+a_4+a_5 +
m_{0,3} + m_{1,3} + m_{2,3} = C$.  Thus $a_2 = a_3 = m_{0,3} = m_{1,3}
= m_{2,3} = 0$.  This means that sub-blocks of types 2 and 3 cannot
occur and that $M$ has the form
\[
	M \;=\;
	\left( \begin{array}{cccccc}
		m_{0,0} & m_{0,1} & m_{0,2} & 0       & 0 & 0\\
		m_{1,0} & m_{1,1} & m_{1,2} & 0       & 0 & 0\\
		m_{2,0} & m_{2,1} & m_{2,2} & 0       & 0 & 0\\
		0	& a       & 0       & C-a     & 0 & 0\\
		n^2	& 0	  & n^2-a   & a       & 0 & 0\\
		0	& 0	  & 0 	    & 0       & 0 & 0
	       \end{array}
	\right)
\]
where we write $a = a_5$, for simplicity.

Projections of row $0$ and column $0$ imply $m_{0,0}\le R$ and
$m_{0,0}+m_{1,0}+m_{2,0} = n^2+R$, and from~(\ref{eq-L1}) we have
$m_{1,0}+m_{2,0} \le n^2$.  Therefore $m_{0,0}=R$ and
$m_{1,0}+m_{2,0}=n^2$. The first equation forces $m_{0,1} = m_{0,2} =
0$, while the second forces $m_{2,1}=0$, by (\ref{eq-L2}).  Hence
\[
	M \;=\;
	\left( \begin{array}{cccccc}
		R       & 0	  & 0 	    & 0       & 0 & 0\\
		b       & m_{1,1} & m_{1,2} & 0       & 0 & 0\\
		n^2-b   & 0       & m_{2,2} & 0       & 0 & 0\\
		0	& a       & 0       & C-a     & 0 & 0\\
		n^2	& 0	  & n^2-a   & a       & 0 & 0\\
		0	& 0	  & 0 	    & 0       & 0 & 0
	       \end{array}
	\right)
\]
for some integer $b\ge0$. Projections of column $1$ and rows $1,2$ 
imply that
\[
	M \;=\;
	\left( \begin{array}{cccccc}
		R       & 0	  & 0 	    & 0       & 0 & 0\\
		b       & Y+C-a   & a-b     & 0       & 0 & 0\\
		n^2-b   & 0       & B+R+b   & 0       & 0 & 0\\
		0	& a       & 0       & C-a     & 0 & 0\\
		n^2	& 0	  & n^2-a   & a       & 0 & 0\\
		0	& 0	  & 0 	    & 0       & 0 & 0
	       \end{array}
	\right).
\]
By inequality~(\ref{eq-L1}) for $(i,j)=(3,1)$ we have $n^2-b+a\le
n^2$, so $a=b$, because all entries are non-negative. Thus $m_{1,2} =
0$.

Write $(i,j) \bowtie (i',j')$ if each block has a center at exactly
one of the positions $(i,j)$ or $(i',j')$. Clearly, if $m_{i,j} +
m_{i',j'} = n^2$ and the tiles centered at $(i,j)$, $(i',j')$ overlap,
then $(i,j)\bowtie (i',j')$. Since $a=b$,
by the form of $M$ above, we get the following relations:
\begin{eqnarray*}
(1,0) \;\bowtie\; (2,0) \;\bowtie\; (3,1) \;\bowtie\; (4,2) \;\bowtie\; (4,3)
		\quad \mbox{\rm and} \quad
(1,1) \;\bowtie\; (2,2).
\end{eqnarray*}
Write $(i,j) \equiv (i',j')$ if each block has 
a tile centered at $(i,j)$ iff it has a tile centered at
$(i',j')$. By the above, we get
\begin{eqnarray*}
(1,0) \;\equiv\; (3,1) \;\equiv\; (4,3)
                \quad \mbox{\rm and} \quad
(2,0) \;\equiv\; (4,2) 
\end{eqnarray*}
By extending the three allowed configurations (number 1,4,5) of the
rows $3,4,5$, and using the above constraints, we obtain
that the only block tilings that meet these conditions are the
admissible tilings in Figure~\ref{Lshape}.
\end{proof}


\section{Conclusion}

We proved {\NP}-completeness for several variants of the tiling problem,
but what can be said about the infinitely many variants for which the
complexity remains open?

A tile $t$ of width $w$ and height $h$ is said to be \emph{interlocking}
if there is a box of width $<2w$ and of height $<2h$ which contains
two disjoint copies of $t$. For example, the tiles from
Theorems~\ref{thm-shape} and \ref{thm-Lshape} are
interlocking, while rectangles are not. There are non-rectangular
tiles that are not interlocking, for example the U-shaped tiles.
For instances consisting of one interlocking tile we believe the 
tiling problem to be {\NP}-complete. 

We are also confident that the problem is {\NP}-complete for all variants
involving at least two different tiles, one of width $\ge2$ and one 
(possibly the same one) of height $\ge 2$. This condition ensures that 
the problem is not invariant under column or row permutations. 

We believe that the techniques developed in this paper will be useful
in developing generic transformations that can show {\NP}-completeness of
wide classes of tiling problems. 

\bibliographystyle{alpha} \bibliography{tomoTiling}

\begin{thebibliography}{GGP00}

\bibitem[CD01]{ChrobakDurr01}
M.~Chrobak and C.~D\"urr.
\newblock Reconstructing polyatomic structures from discrete {X}-rays:
  {NP}-completeness proof for three atoms.
\newblock {\em Theoretical Computer Science}, 259:81--98, 2001.

\bibitem[DGRR]{DGRR00}
Christoph D\"urr, Eric Goles, Ivan Rapaport, and Eric Remila.
\newblock Tiling with bars under tomographic constraints.
\newblock {\em Theoretical Computer Science}.
\newblock to appear.

\bibitem[GGP00]{GaGrPr00}
R.~Gardner, P.~Gritzmann, and D.~Prangenberg.
\newblock On the computational complexity of determining polyatomic structures
  by {X}-rays.
\newblock {\em Theoretical Computer Science}, 233:91--106, 2000.

\bibitem[GLS88]{GrLoSc}
M.~Gr\"otschel, L.~Lov\'asz, and A.~Schrijver.
\newblock {\em Geometric algorithms and combinatorial optimization}.
\newblock Springer-Verlag, 1988.

\bibitem[Pic01]{Picouleau99}
Christophe Picouleau.
\newblock Reconstruction of domino tiling from its two orthogonal projections.
\newblock {\em Theoretical Computer Science}, 255:437--447, 2001.

\bibitem[Rys57]{Ryser57}
H.J. Ryser.
\newblock Combinatorial properties of matrices of zeroes and ones.
\newblock {\em Canad. J. Math.}, 9:371--377, 1957.

\bibitem[Rys63]{Ryser63}
H.J. Ryser.
\newblock {\em Combinatorial Mathematics}.
\newblock Mathematical Association of America and Quinn \& Boden, Rahway, New
  Jersey, 1963.

\end{thebibliography}
\end{document}